\documentclass{aastex}
\usepackage{emulateapj5,apjfonts,epsfig}

\slugcomment{Accepted for publication in The Astrophysical Journal, 2001 Aug 3}
\shorttitle{MILLIHERTZ OSCILLATIONS IN 4U 1626--67}
\shortauthors{CHAKRABARTY ET. AL.}

\begin{document}

\title{Millihertz Optical/UV Oscillations in 4U~1626--67: Evidence for
a Warped Accretion Disk\altaffilmark{1}} 

\author{Deepto~Chakrabarty}
\affil{\footnotesize Department of Physics and Center for Space Research, 
   Massachusetts Institute of Technology, Cambridge, MA 02139}
\email{deepto@space.mit.edu}

\author{Lee Homer\altaffilmark{2} and Philip~A.~Charles\altaffilmark{3}}
\affil{\footnotesize Nuclear and Astrophysics Laboratory, University of
   Oxford, Keble Road, Oxford, OX1 3RH, UK}
\email{homer@astro.washington.edu, pac@astro.soton.ac.uk}

\and

\author{Darragh O'Donoghue}
\affil{\footnotesize South African Astronomical Observatory, P.O. Box 9, 
   Observatory 7935, Cape Town, South Africa}
\email{dod@da.saao.ac.za}

\altaffiltext{1}{Based in part on observations with the
NASA/ESA {\em Hubble Space Telescope}, obtained at the Space Telescope
Science Institute, which is operated by the Association of
Universities for Research in Astronomy, Inc., under NASA contract NAS
5-26555.}

\altaffiltext{2}{Current address: Department of Astronomy, University
of Washington, Box 351580, Seattle, WA 98195.}

\altaffiltext{3}{Current address: Department of Physics and Astronomy,
   University of Southampton, Southampton SO17 1BJ, UK.}

\begin{abstract}
We have detected large-amplitude 0.3--1.2~mHz quasi-periodic
oscillations (QPOs) from the low-mass X-ray binary pulsar
4U~1626--67/KZ TrA, using ultraviolet photometry from the {\em Hubble
Space Telescope} and ground-based optical photometry.  These 1 mHz
QPOs, which have coherence ($\nu/\Delta\nu)\approx 8$, are entirely
distinct from the 130 mHz pulsar spin frequency, a previously known 48
mHz QPO, and the 42 min binary period (independently confirmed here).
Unlike the 48 mHz and 130 mHz oscillations which are present in both
the optical/UV and the X-ray emission, the 1 mHz QPOs are not detected
in simultaneous observations with the {\em Rossi X-Ray Timing
Explorer}.  The rms amplitude of the mHz QPO decreases from 15\% in
the far UV to 3\% in the optical, while the upper limit on a
corresponding X-ray QPO is as low as $<0.8$\%.  We suggest that the
mHz oscillations are due to warping of the inner accretion disk.  We
also report the detection of coherent upper and lower sidebands of the
130 mHz optical pulsations, with unequal amplitude and a spacing of
1.93 mHz around the main pulsation.  The origin of these sidebands remains
unclear. 
\end{abstract}


\section{INTRODUCTION}

The optical and ultraviolet emission from low-mass X-ray binaries
(LMXBs) is primarily powered by irradiation of the accretion disk
and/or the mass donor by the central X-ray source.  In short period
binaries, this reprocessed emission generally dominates both the
internal (viscous) heating in the disk (except perhaps at the smallest
disk radii) and the intrinsic luminosity of the donor star (see van
Paradijs \& McClintock 1995 for a review).  Since the time history of
the optical/UV emission should thus be a convolution of the X-ray
intensity history with a function representing the binary and disk
geometry, we expect that periodic or quasi-periodic modulation of the
optical/UV emission should arise from one of only two causes: (1) a
similarly modulated ``input'' X-ray intensity, as in coherent X-ray
pulsations or quasi-periodic oscillations (QPOs) (see, e.g.,
Chakrabarty 1998); and (2) orbitally-modulated X-ray heating of the
binary companion's surface (see van Paradijs \& McClintock 1995).  In
this paper, we present evidence that a third mechanism must exist, and
suggest implications for the smoothness of the accretion disk surface
in LMXBs.

The LMXB 4U 1626--67 consists of a 7.66 s X-ray pulsar
accreting from an extremely low-mass ($\lesssim 0.1 M_\odot$)
companion in an ultracompact 42 min orbit (Middleditch et al. 1981;
Levine et al. 1988; Chakrabarty 1998).  The 130 mHz (=1/7.66 s) X-ray
pulsations arise from anisotropic accretion of matter on the surface of a
rotating, highly-magnetized neutron star whose spin and magnetic
dipole axes are misaligned.  A 48 mHz QPO is detected in the X-ray
emission (Shinoda et al. 1990; Kommers, Chakrabarty, \& Lewin 1998),
and probably arises via an interaction between the pulsar's
magnetosphere and the accretion disk.  

The optical counterpart, KZ Triangulum Australis (KZ TrA), has a
strong ultraviolet excess (McClintock et al. 1977), and optical
pulsations are detected at the same 130 mHz frequency as the X-ray
pulsations (Ilovaisky et al. 1978).  The pulsed optical emission from
the system is understood as primarily due to X-ray reprocessing in the
accretion disk (Chester 1979; McClintock et al. 1980; Chakrabarty
1998).  The power spectral peak of the optical pulsations has a weak
but persistent lower-frequency sideband ($\Delta\nu=$0.4 mHz)
which has been interpreted as due to X-ray reprocessing on the surface
of a companion star in a 42-min prograde binary orbit (Middleditch et
al. 1981; Chakrabarty 1998).  In addition, the 48 mHz QPO detected in
the X-ray emission is also detected in the optical band (Chakrabarty
1998).  During the 1970s and 1980s, the system also showed strong,
correlated X-ray/optical flares every $\sim$1000 s that are of
undetermined origin (Joss, Avni, \& Rappaport 1978; McClintock et
al. 1980; Li et al. 1980).

\begin{figure*}[t]
\centerline{\epsfig{file=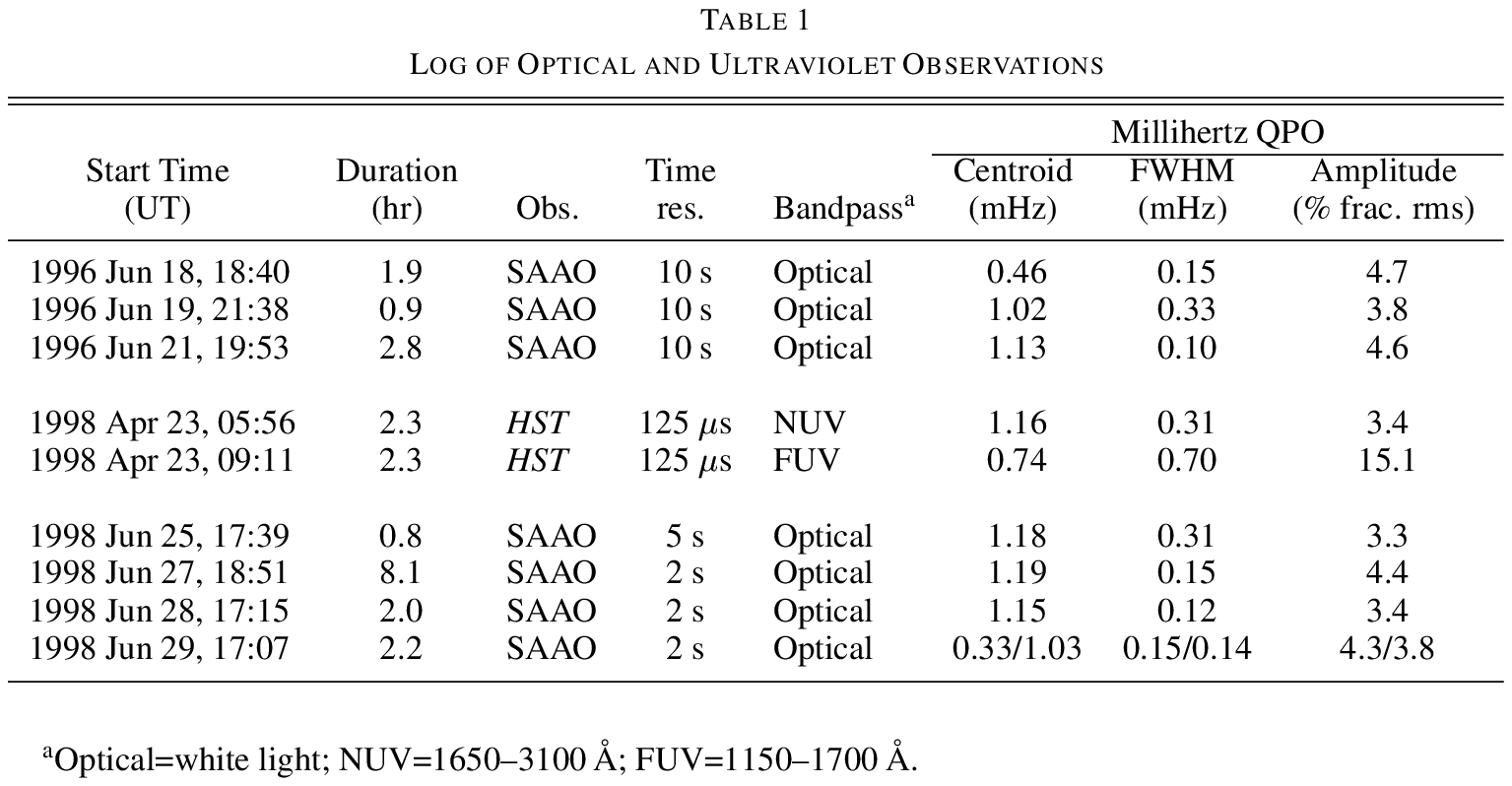}}
\end{figure*}
Overall, the known time variability of the optical emission from 4U
1626--67 is closely correlated with the X-ray emission.  In this paper,
however, we report on the discovery of two previously unknown timing
features in the emission from 4U 1626--67:  a strong 0.3--1.2~mHz
quasi-periodic modulation in the optical and ultraviolet emission, and
lower and upper sidebands of the coherent pulsations with symmetric
separation from the main peak ($\Delta\nu=1.93$ mHz) but asymmetric
strengths.  These new modulations are absent in simultaneous X-ray
data.   

\section{OBSERVATIONS}

\subsection{Ultraviolet observations}

Our {\em Hubble Space Telescope (HST)} observations of KZ TrA were
made on 1998 April 23 using the Space Telescope Imaging Spectrograph
(STIS; Kimble et al. 1998).  A log of the observations is given in
Table 1.  Two {\em HST} orbits ($P=96$ min) of data were acquired
using the Cs$_2$Te multi-anode microchannel array (MAMA) detector
(STIS/NUV-MAMA), which operates in the near ultraviolet band (NUV;
1650--3100~\AA).  Another two orbits of data were acquired using the
solar-blind CsI MAMA (STIS/FUV-MAMA), which operates in the far
ultraviolet band (FUV; 1150--1700~\AA).  Useful data were obtained for
about 45 min of each 96 min {\em HST} orbit.  The remainder of each
orbit was occupied by guide star acquisition, instrumental calibration
and readout overheads, and Earth occultations of the source.  

Our observations were acquired in a backup guiding mode using only a
single guide star, because one of the {\em HST} Fine Guidance Sensors
was unable to achieve fine lock on its guide star.  The telescope
pitch/yaw angles were controlled using the one available guide star,
and the roll angle by gyroscope.  Both the STIS initial target
acquisition and 5-point linear peakup procedure succeeded normally.
The small drifts in roll angle that might have occurred in this
gyroscope mode could not give rise to a mHz QPO signal of the strength
that we are reporting in this paper (and which are also seen
independently in our ground-based optical data\footnote{Indeed,
additional {\em HST}/STIS data of the source taken in 1999 also contain
these mHz oscillations.}).

Both STIS MAMA cameras were operated in TIMETAG mode, which records
both the detector coordinate and the arrival time ($\Delta t=16 \mu$s)
of each detected photon individually.  The observations were made
through a 52$\times$0.2 arcsec$^2$ long slit and a low-resolution
grating (G230L for the NUV observation and G140L for the FUV
observation) in order to obtain spectral information as well.
However, for the analysis presented here, we summed the spectra along
the dispersion axis in order to obtain broad-band photometry over the
entire bandpass of each camera.  The complete NUV and FUV light curves
are shown in the top two panels of Figure~1.  A spectroscopic analysis
will be reported separately (Wang \& Chakrabarty 2001).

\subsection{Optical observations}

We observed a small (50$\times$33 arcsec$^2$) region surrounding
KZ TrA using the UCT-CCD fast photometer (O'Donoghue 1995) at the
Cassegrain focus of the 1.9-m telescope at the South African
Astronomical Observatory (SAAO) in Sutherland, South Africa, on 1996 June 
19--22 and on 1998 June 25--30.  A detailed observation log is given
in Table~1.  The UCT-CCD fast photometer is a Wright Camera
576$\times$420 coated GEC CCD, which was used half-masked and operated
in frame-transfer mode, allowing exposures as short as 1 s with no
dead time.  Unlike photoelectric aperture photometry, the CCD images
automatically provide a simultaneous sky background measurement over
the entire observation.  

The 1996 observations were taken at 10~s resolution, and the 1998
observations were taken at either 2~s or 5~s resolution.  All the
observations were taken in white light with no bandpass filters in
place.  The data reduction procedure is described by Homer, Charles,
\& O'Donoghue (1998).  A preliminary analysis of the 1996 observations
was presented by Homer (1998).  An excerpt of the light curve from the
1998 June 29 observation is shown in the bottom panel of Figure~1.

\subsection{X-ray observations}

For both the 1998 April {\em HST} and the 1998 June SAAO
observations, coordinated simultaneous X-ray observations were made 
using the {\em Rossi X-Ray Timing Explorer (RXTE)}.  A log of these
observations is given in Table~2.  The analysis presented here uses
data from the {\em RXTE} Proportional Counter
\centerline{\epsfig{file=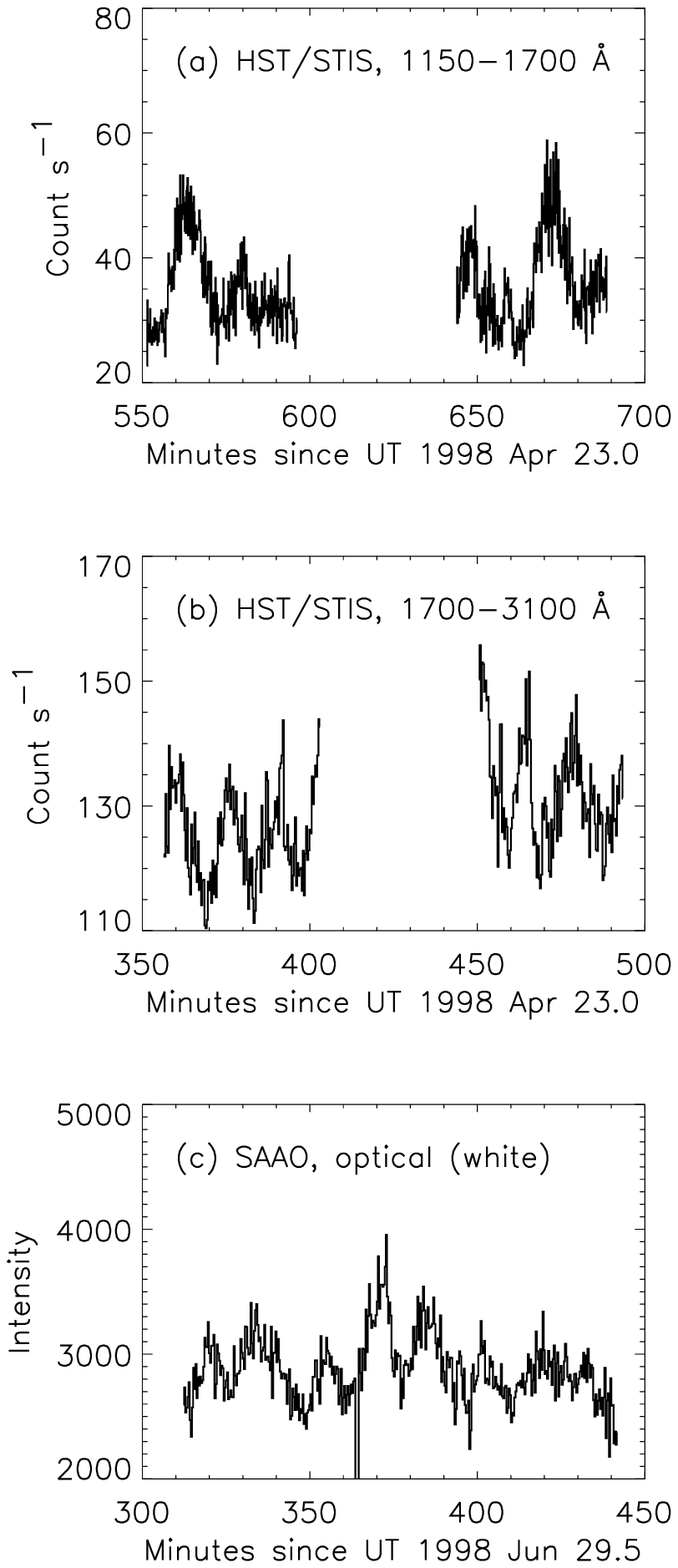,width=0.9\linewidth}}
\figcaption{UV and optical light curves of 4U 1626--67/KZ TrA.  (a)
Far-UV (1150--1700 \AA) {\em HST}/STIS light curve from 1998 Apr
23. (b) Nearr-UV (1700--3100 \AA) {\em HST}/STIS light curve from 1998
Apr 23.  (c) Optical (white light) light curve from 1998 June 29.}
\vspace*{0.1in}
\noindent
Array (PCA; Jahoda et al. 1996), which operates in the 2--60 keV range
with a total effective area of $\sim$6500 cm$^2$ and a collimated
1$^\circ$ field of view.  The PCA data were recorded in a variety of
different data modes simultaneously.  For our analysis, we used the
{\tt Standard1} data (0.125 s time resolution, no energy resolution)
and the {\tt Standard2} data (16 s time resolution, 128 channel energy
resolution).

During the 1998 April observations, 4U 1626--67 was in the continuous
viewing zone (near the orbital pole) of {\em RXTE}, so that it was
never subject to Earth occultation.  This allowed unusually long
uninterrupted observations.  The short gaps between these observations
were due to spacecraft passages through the South Atlantic Anomaly
(SAA; a region of extremely high 
\centerline{\epsfig{file=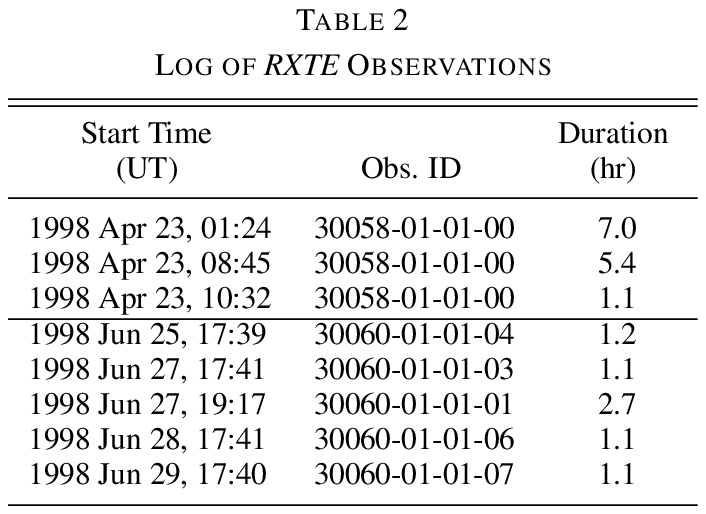,width=0.9\linewidth}}
\break
\noindent
particle background in the Earth's
magnetosphere) during which the detector high-voltage was turned off.
The 1998 June observations were subject to both Earth occultations and
SAA passages, and thus contain gaps. 

\section{ANALYSIS AND RESULTS}

There were a variety of oscillations present in our data, both
quasi-periodic and periodic, some previously known and some reported
here for the first time.   In order to make a quantitative analysis of
the timing properties of these data, we computed the Fourier
power spectrum for each observation.   We adopted the
root-mean-squared (rms) normalization convention widely used in the
QPO literature (e.g., Miyamoto et al. 1994; van der Klis 1995).
Unlike simple photon-counting data, the measurement noise in our data
contains additional contributions beyond counting statistics.
Nevertheless, we found that the overall measurement noise in our CCD
data is well-described by a Gaussian white noise process, and we
simply estimated the white noise level from the high-frequency
($>100$~mHz) power spectrum of the data and subtracted it off, as with
the Poisson noise case for photon counting data.  Table~3 summarizes
the amplitudes of the 1 mHz and 48 mHz QPOs, and Table~4 summarizes
the pulsed amplitude of the coherent pulsations.  

\subsection{1 mHz QPO}

All the optical and ultraviolet data sets showed a clear $\sim$1 mHz
modulation in intensity.  There is no previously known QPO at this
frequency from 4U 1626--67, although bright X-ray and optical flaring
on 1000~s time scales has been seen (see \S4).  As shown in Figure~1, 
individual cycles of this modulation are clearly visible in the data.  The
centroid frequency of the modulation was always around 1 mHz, although
values as low as 0.46 mHz and as high as 1.19 mHz were observed.  On
1998 June 29, a pair of distinct modulations of comparable amplitude
was detected with centroid frequencies of 0.33 mHz and 1.03 mHz.  A
summary of the measured QPO frequencies, widths, and amplitudes is
given in Table~1.  

In the longest observations, the 1 mHz modulation was resolved in the
power spectrum and had a coherence $Q\equiv\nu/\Delta\nu$ in the range
of 8--11.  A power spectrum of the longest optical observation is
shown in Figure~2.  The limited frequency resolution of the shorter
observations yielded only lower limits for $Q$ which are consistent
with this range.  The root-mean-squared (rms) fractional amplitude of
the modulation was in the 3--5\% range for the optical and NUV
observations and over 15\% for the FUV data.
\centerline{\epsfig{file=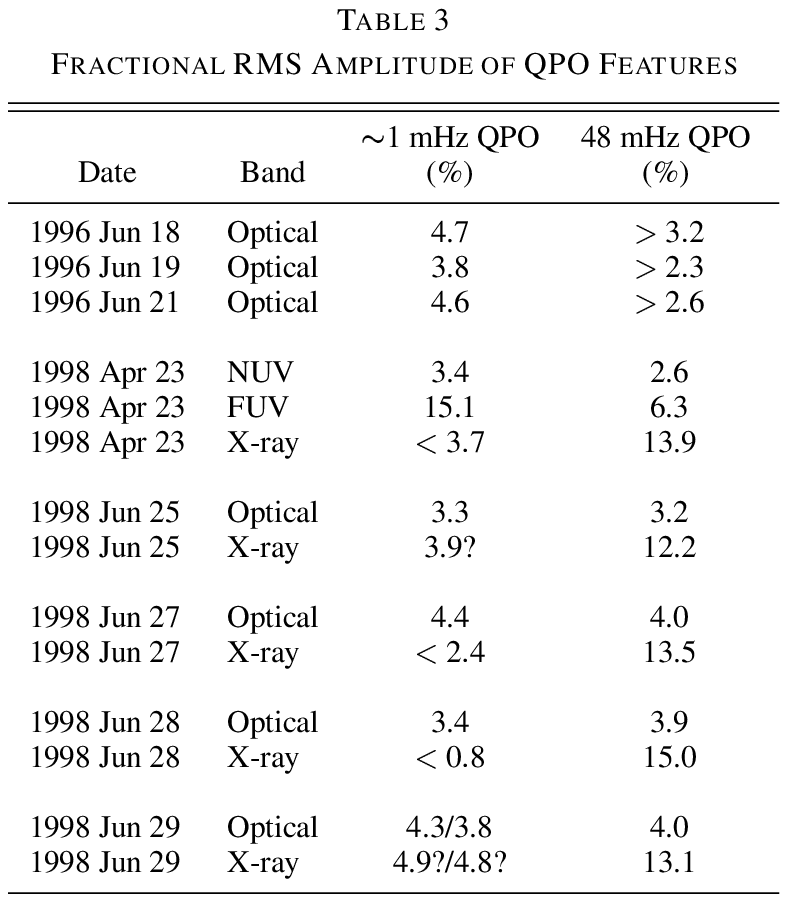,width=0.9\linewidth}}
\break

No 1 mHz modulation was detected in simultaneous X-ray observations in
1998 April and June, with 3$\sigma$ upper limits ranging from $<0.8$\%
to $<5$\%.   A power spectrum of the longest X-ray observation (which
provided the most stringent upper limit) is shown in Figure~3.   A
comparison of the mHz QPO amplitude in the different bands is given
in Table~3.  

\subsection{48 mHz QPO}

A 48 mHz QPO was detected in all of our optical, ultraviolet, and
X-ray data, and is clearly visible in Figure 2 and 3. This feature has
been previously detected at both X-ray and optical wavelengths
(Shinoda et al. 1990; Chakrabarty 1998).  Its centroid and width
($Q\approx 4$) were both quite stable across all our observations.
In the ultraviolet data, individual cycles of the QPO were visible.
The fractional rms amplitude of the X-ray QPO was in the 12--15\%
range, while the amplitude of the simultaneous optical QPO was 3--4\%.
The optical/X-ray amplitude ratio for this QPO had a stable value of
$\approx 30\%$ over all the simultaneous observations.   The NUV/X-ray
and FUV/X-ray amplitude ratios were significantly lower and higher,
respectively.   A summary of the 48 mHz QPO amplitude in the various
observations is given in Table~3.

\subsection{130 mHz pulsation and related features}

A coherent pulsation near 130.4 mHz, corresponding to the neutron star
spin frequency (which evolves on a time scale of $|\nu/\dot\nu|\approx
5000$~yr; Chakrabarty et al. 1997), was detected in all of the X-ray
and ultraviolet observations, as well as every optical observation
with sufficiently fast sampling.  This optical pulsation has been well
studied previously and was also accompanied by the usual symmetric sidelobe
structure ($[\sin^2 \pi\nu/(\pi\nu)^2]$; see, e.g., van der Klis 1989)
due to the finite data length.  In addition, the 260.8 mHz second
harmonic\footnote{We have adopted the convention that $m\nu$ is the
frequency of the $m$th harmonic for a fundamental frequency $\nu$.}
\centerline{\epsfig{file=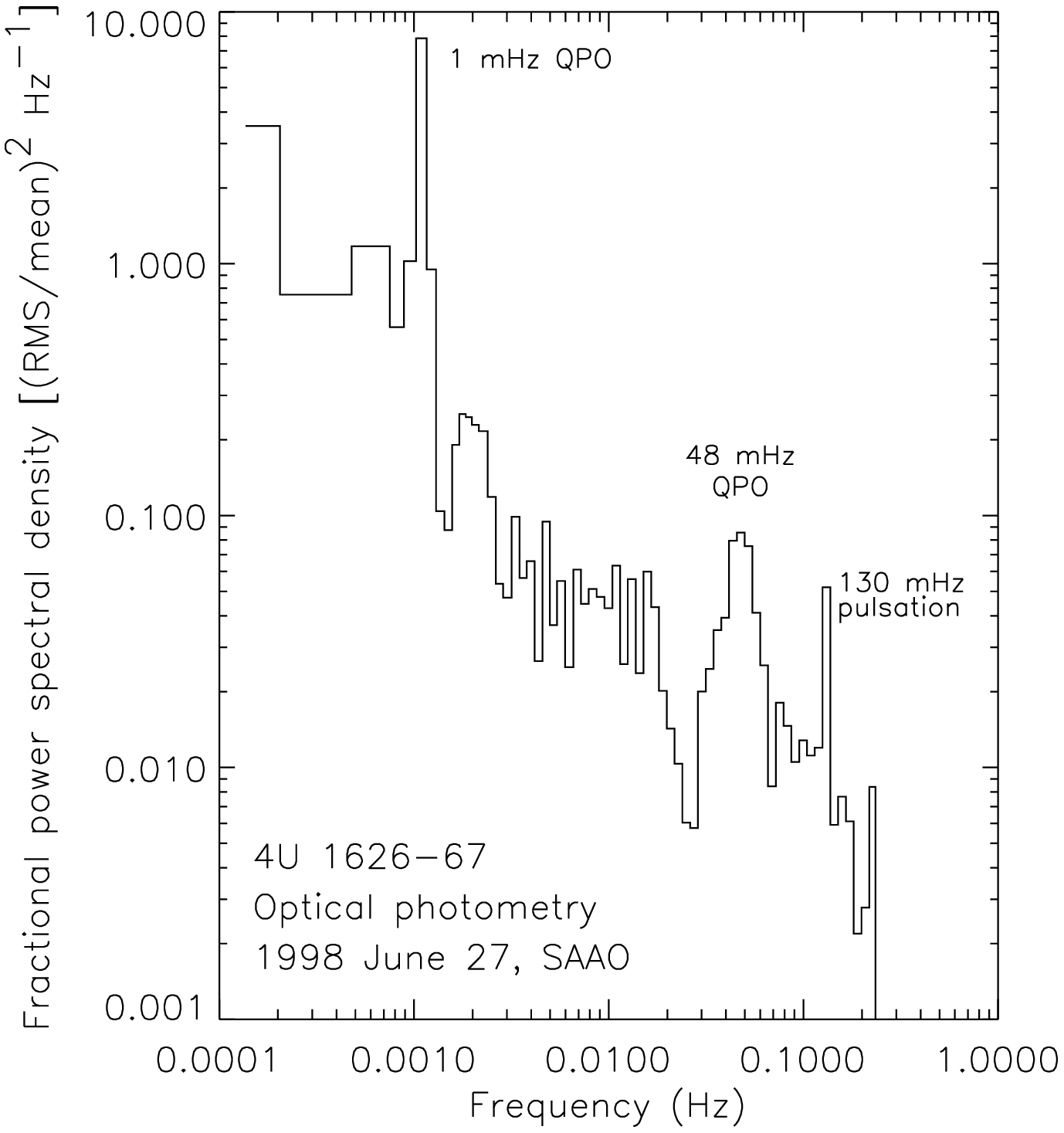,width=0.9\linewidth}}
\figcaption{Fourier power density spectrum of an 8.1-hr white light
observation of 4U 1626--67 on 1998 June 27, normalized relative to the
mean source power.  A 1 mHz QPO, as well as a previously known 48 mHz
QPO and 130 mHz pulsation, are visible in the data.}
\vspace*{0.1in}
\centerline{\epsfig{file=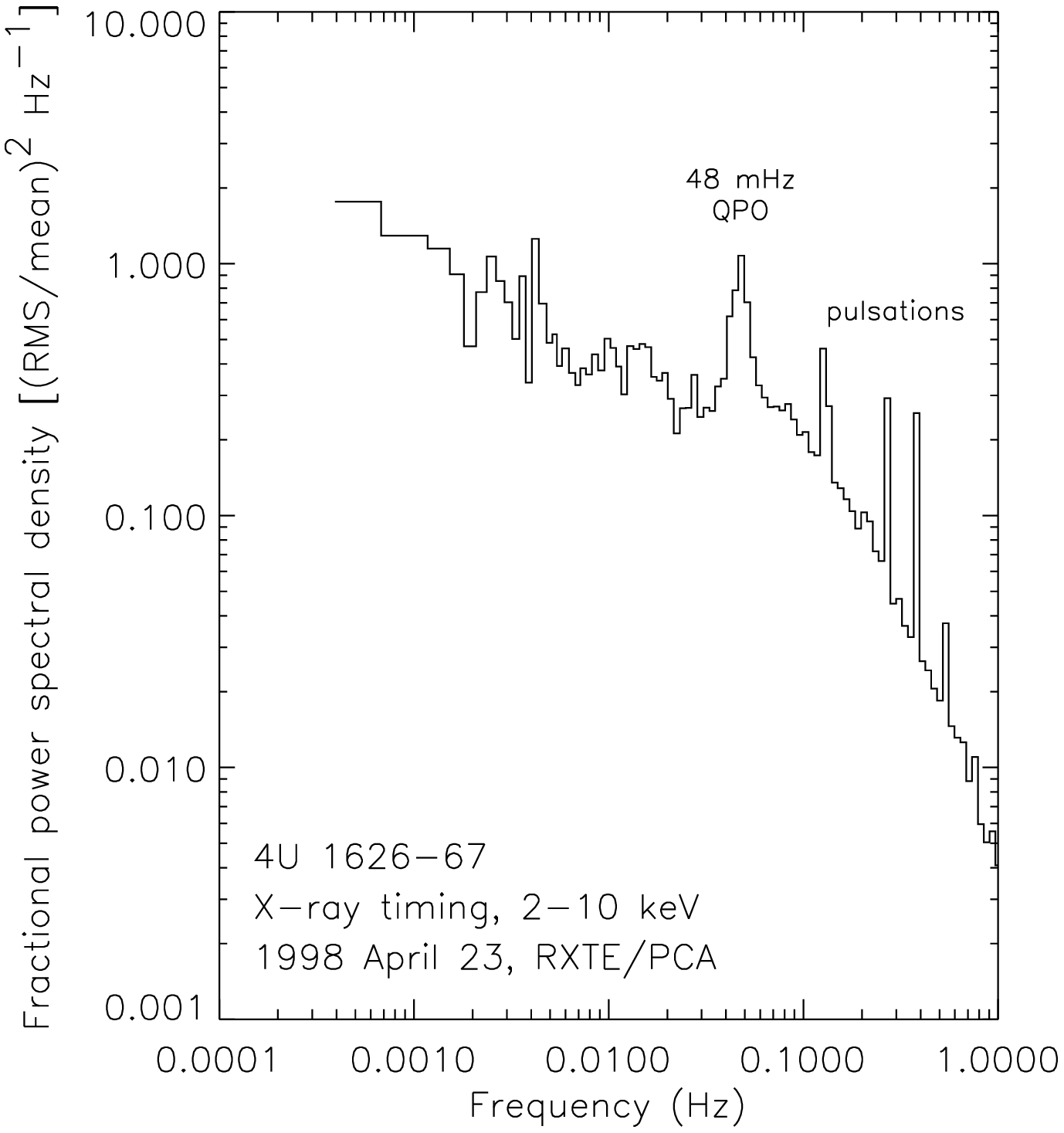,width=0.9\linewidth}}
\figcaption{Fourier power density spectrum of an 2.3-hr 2--10 keV {\em
RXTE}/PCA observation of 4U 1626--67 on 1998 Apr 23, normalized
relative to the mean source power.  This observation is simultaneous
with the {\em HST} UV observations shown in Figure 1, both of which show
a 1 mHz QPO clearly.  The 1~mHz QPO is undetected in the X-ray data,
although the previously known 48 mHz QPO and several harmonics of the
130 mHz pulsation are clearly visible.}
\vspace*{0.1in}
\noindent
was also detected in all of the X-ray data and some of the optical/UV
observations.  Note that this frequency slightly exceeds the 250 mHz
Nyquist frequency for the 2~s optical data, causing the harmonic to
actually be detected at an aliased frequency $\nu_{\rm alias} =
\nu_{\rm Nyquist} - (\nu\,\mbox{\rm mod}\,\nu_{\rm Nyquist})=239.2$~mHz.
Binning effects cause aliased signal amplitudes in this region to be
suppressed by a factor of 0.61 relative to the true signal amplitude
(see, e.g., Leahy et al. 1983).
\begin{figure*}[t]
\centerline{\epsfig{file=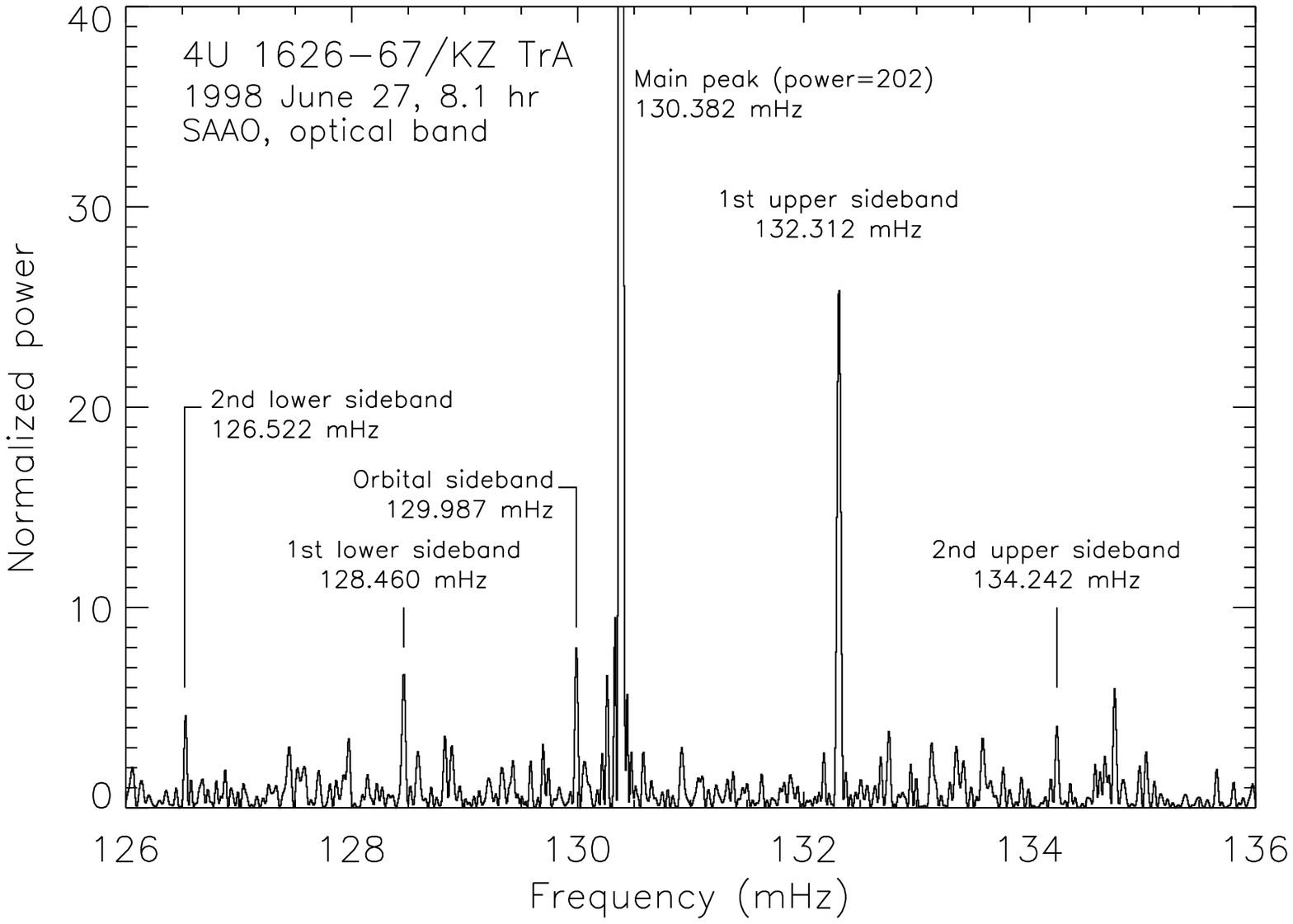,width=0.9\linewidth}}
\figcaption{Expanded view of an oversampled unit-normalized Fourier
power spectrum of the 8.1 hr white light observation of 4U 1626--67 on
1998 June 27, centered on the 130.382 mHz pulsation.  The previously
known orbital lower sideband ($\Delta\nu=0.395$ mHz) is shown, as well as
two pairs of previously unknown equispaced lower and upper sidebands
with $\Delta\nu=1.93$ mHz.}
\end{figure*}

In several of the optical/UV observations, the 130.4 mHz pulsation is
accompanied by a lower frequency sideband with a separation of
0.395(5) mHz.  This feature, which was first reported by Middleditch
et al. (1981) and later confirmed by Chakrabarty (1998), is understood
as reprocessing from the surface of the mass donor star.  The downward
shift relative to the main pulsation frequency implies a prograde 42
min binary orbit for the system.  This is the first time that the
orbital feature has been confirmed by an instrument other than the
ASCAP single-channel photometer at the Cerro Tololo Inter-American
Observatory.

We computed an oversampled power spectrum of the longest optical
observation (1998 June 27) in order to study the timing properties of
the coherent pulsations in more detail.  The oversampled power
spectrum in this frequency region is shown in Figure 4.  Besides the
main pulsation [130.382(1) mHz] and the orbital sideband [129.987(5) mHz],
two additional pairs of sidebands are also detected.  These sidebands are
identically spaced around the main pulsation at frequencies $\nu_{\rm
spin}\pm\Delta\nu$ and $\nu_{\rm spin}\pm 2\Delta\nu$ with 
$\Delta\nu=1.93(1)$ mHz.  In the first sideband pair, the rms
amplitude of the upper sideband is twice that of the lower sideband,
but the second pair of sidebands have equal strength.  A single pair of
sidebands with the same frequency separation are also detected around
the (aliased) second harmonic of the main pulsation.  These sidebands
are also asymmetric in amplitude, but it is the (apparent) lower
sideband which is stronger than the upper sideband.   However, due to
aliasing,  the (true) higher frequency sideband actually appears as a
{\em lower} sideband, and vice versa, so the sense of the amplitude
asymmetry is in fact the same as in the first sidebands of the
fundamental. These 1.93~ mHz sideband pairs are not detected in any of
the other optical, UV, or X-ray observations\footnote{Sideband pairs
with unequal amplitudes have been previously observed around the X-ray
pulsation (Kommers et al. 1998).  However, these sidebands are spaced
by the 48 mHz QPO frequency.} However, we note that the other
optical/UV observations were of shorter duration (and generally less
sensitive), while the X-ray observations were measuring direct rather
than reprocessed emission. 

A summary of the rms amplitudes of the coherent pulsations and
associated sidebands in each of the various observations is given in
Table~4.  
\begin{figure*}[t]
\centerline{\epsfig{file=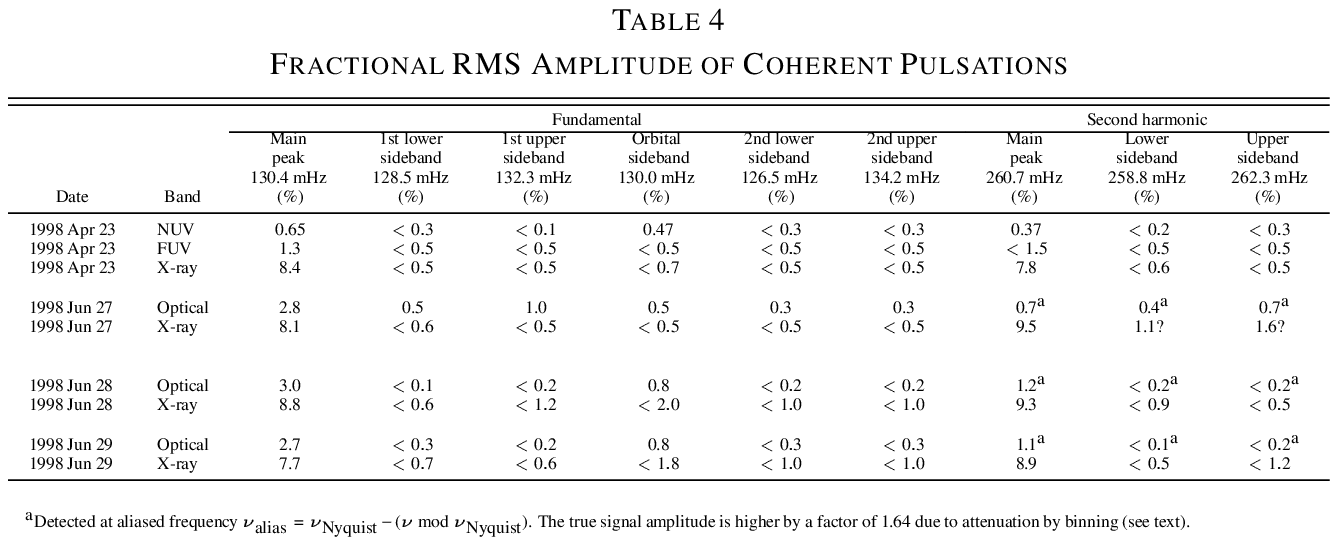,width=0.9\linewidth}}
\end{figure*}

\section{DISCUSSION}

We have identified several persistent, distinct quasi-periodic and
periodic oscillations in the optical and ultraviolet flux from
4U~1626--67.  Some of these oscillations (the 130 mHz pulsation and
its harmonic, as well as the 48 mHz QPO) are also present in
simultaneous X-ray data.  This is consistent with the optical/UV
variability arising from the reprocessing of variable X-ray emission
(which itself arises from accretion onto the pulsar) in the accretion
disk, as recognized previously (Chester 1979; McClintock et al. 1980;
Chakrabarty 1998).  The 0.395 mHz orbital lower sideband of the 130
mHz optical pulsation has no X-ray counterpart, but is nevertheless
understood as reprocessing from the surface of the pulsar's binary
companion (Middleditch et al. 1981).  

However, the 1 mHz QPO, while strong in the optical/UV emission, is
undetected in simultaneous X-ray data, and thus cannot easily be
attributed to reprocessing of variable X-ray illumination.  The
feature is fairly narrow in frequency, unlike the broader mHz
ultraviolet QPOs recently reported in the LMXB pulsar Her X-1 by
Boroson et al. (2000).  The time scale of the 1~mHz QPO is similar to
the flaring time-scale seen in simultaneous X-ray and optical data
during the 1970s and 1980s (Joss et al. 1978; McClintock et al. 1980;
Li et al. 1980).  However, those events were abrupt, well-defined
flares occuring on $\sim$1000~s time scales and lasting $\sim$500~s,
with flux enhancements of (1.3--3)$\times$ the quiescent level; they
were also simultaneously observed in both the X-ray and optical
bands. This is quite unlike the quasi-sinusoidal optical/UV
modulations we report here.  Interestingly, the $\sim$1000~s
X-ray/optical flares have not been seen in any data acquired since
1990 (Chakrabarty et al. 2001, in preparation), possibly related to
the abrupt 1990 change in the pulsar's accretion torque state and
X-ray spectrum (Chakrabarty et al. 1997).  In any case, the optical
flares were consistent with reprocessing of X-ray flares by the
accretion disk.

We suggest that the 1~mHz QPO modulation arises from a purely geometric
effect: a ``warping'' of the accretion disk, wherein the tilt angle of
the normal to the local disk surface varies with azimuth.  In this
case, we would expect the flux of reprocessed emission visible along
our line of sight to vary as the reprocessing geometry evolves due to
the warp.  We note that the 1~mHz QPO is strongest in the FUV band;
emission in this band most likely originates in the inner accretion
disk, which is truncated by the pulsar magnetosphere near the pulsar's
corotation radius $r_{\rm co}=6.5\times 10^8$~cm (see Chakrabarty
1998).  The weaker amplitude of the QPO in the optical band suggests
that the warp primarily affects the inner disk.  As such, we might
expect this phenomenon to be difficult to detect in high-inclination
binaries, where shadowing by the outer disk may be a strong effect.
It is thus interesting to note that 4U~1626--67 is a very
low-inclination (nearly face-on) binary (e.g., Levine et al. 1988;
Chakrabarty 1998).  A test of our warped disk hypothesis could be made by
observing the 1~mHz QPO simultaneously in the UV and the optical,
since this ought to allow a measurement of the geometry of the warp
over the entire disk.  

The possibility of accretion disk warping has attracted considerable 
recent theoretical interest.  Several authors have shown that if an
accretion disk is subject to sufficiently strong central irradiation, 
then it is unstable to warping (Petterson 1977; Pringle 1996; Maloney
et al. 1996).  Pringle (1996) showed that even an initially flat disk
is unstable to warping, and Wijers \& Pringle (1999) suggested that
warping should be generic in irradiated disks, but Ogilvie \&
Dubus (2001) argue that warping will only occur under a limited range
of conditions.  We note, however, that in the Pringle instability the
warp is strongest in the outer disk and has time scales of order days
to weeks; it is therefore unclear whether this model can explain our
mHz QPO.  An alternative disk warping mechanism based on magnetic
torques acting on the inner disk has also been proposed, and may be
the disk-magnetosphere interaction in the source (Lai 1999; Terquem \&
Papaloizou 2000).

The origin of the coherent $\Delta\nu=1.93$ mHz optical sidebands is
unclear.  Although the separation frequency is of the same order as
the frequency of the 1~mHz QPO, the sidebands are very narrow in
frequency, unlike the QPO.  Warner (1986, 1995) has shown that optical
reprocessing in intermediate polars (accreting magnetic white dwarfs,
analogous to the X-ray pulsars) can give rise to pairs of sidebands
around the spin frequency, spaced by the binary orbital frequency.  It
is possible that 518~s (1/1.93 mHz) is the orbital period, rather than
the previously inferred 42 min ($\approx 2500$ s).  However, we regard
this as unlikely for three reasons.  First, the long-term stability of
the 0.4 mHz lower sideband (Middleditch et al. 1981; Chakrabarty 1998)
is strong evidence for a 42 min orbit.  Second, the optical/UV
spectrum of 4U 1626--67 is best fit with an X-ray heated accretion
disk whose outer radius is $\approx 2\times 10^{10}$ cm (Wang \&
Chakrabarty 2001), which is consistent with a 42 min binary but too
large for a 518~s binary.  Finally, a 518~s binary is likely below the
minimum orbital period ($\sim$10 min) possible for a binary with a
hydrogen-depleted secondary (analogous to the 80 min minimum for
hydrogen-rich secondaries; see Paczynski \& Sienkiewicz 1981 and
Nelson, Rappaport, \& Joss 1986).  

In some ways, the 1.93 mHz sidebands are reminiscent of the
``superhump'' phenomena observed in the SU UMa class of dwarf novae
(see Osaki 1996 and references therein).  During infrequent
``superoutbursts'', these sources show periodic photometric
variability at periods a few percent longer than the binary period.
This variability has been interpreted in terms of a tidally-driven
eccentric instability in the accretion disk, first proposed by
Whitehurst (1988).  This instability evidently requires extreme mass
ratios (i.e., a very low-mass donor, as in 4U 1626--67).  In this
picture, the superhump period arises from a beat between the binary
period and the the slow precession frequency of an eccentric disk. 
Although it is difficult to reconcile this scenario with the 42-min
binary orbit inferred for 4U 1626--67, it is possible that 1.93 mHz
represents a precession frequency in the accretion disk that is
modulating the optical reprocessing.   Alternatively, the sidebands
may arise from reprocessing in some sort of asymmetric feature
orbiting in the disk.  The 1.93~mHz separation corresponds to a
Keplerian radius of $\approx 1\times 10^{10}$ cm, roughly one-third
the binary separation.  Such an asymmetry might be related to warping
of the disk as well. 

\acknowledgements{This paper is based in part on observations made at
the South African Astronomical Observatory (SAAO).  We are grateful to 
Fran\c{c}ois van Wyck and Fred Marang for their support at the telescope.
We also thank our contact scientists at STScI, Stephen Hurlbert and
Michael Asbury, and Evan Smith of the {\em RXTE} Science Operations
Facility, for their assistance in scheduling a simultaneous {\em
HST/RXTE} observation.  This paper was begun during DC's visit to
Oxford as part of the MIT-Balliol College faculty exchange program; DC
thanks Balliol College and his host, Katherine Blundell, for their
hospitality and support.  This work was also supported in part by
the Space Telescope Science Institute under grant GO-7510 and the
NASA {\em RXTE} Guest Observer Program under grant NAG 5-7328.}


\end{document}